# Magnitudes and Timescales of Total Solar Irradiance Variability

Greg Kopp[1],*

1. University of Colorado, Laboratory for Atmospheric and Space Physics, 3665 Discovery Drive, Boulder, CO 80303, USA

*Corresponding author: Greg.Kopp@LASP.Colorado.edu

ABSTRACT

The Sun's net radiative output varies on timescales of minutes to gigayears. Direct measurements of the total solar irradiance (TSI) show changes in the spatially- and spectrally-integrated radiant energy on timescales as short as minutes to as long as a solar cycle. Variations of ~0.01 % over a few minutes are caused by the ever-present superposition of convection and oscillations with very large solar flares on rare occasion causing slightly-larger measureable signals. On timescales of days to weeks, changing photospheric magnetic activity affects solar brightness at the ~0.1 % level. The 11-year solar cycle shows variations of comparable magnitude with irradiances peaking near solar maximum. Secular variations are more difficult to discern, being limited by instrument stability and the relatively short duration of the space-borne record. Historical reconstructions of the Sun's irradiance based on indicators of solar-surface magnetic activity, such as sunspots, faculae, and cosmogenic isotope records, suggest solar brightness changes over decades to millennia, although the magnitudes of these variations have high uncertainties due to the indirect historical records on which they rely. Stellar evolution affects yet longer timescales and is responsible for the greatest solar variabilities. In this manuscript I summarize the Sun's variability magnitudes over different temporal regimes and discuss the irradiance record's relevance for solar and climate studies as well as for detections of exo-solar planets transiting Sun-like stars.

Key words. Total irradiance – Sun – Variability – Climate – Solar activity

## 1. Introduction

The Sun's radiative output provides 99.96 % of the energy driving Earth's climate (Kren 2015). Historically misnamed the "solar constant," the Sun's total radiant energy incident on the Earth varies with time. Even small changes in this energy over long periods of time can affect Earth's climate, as demonstrated in modern times by Eddy (1976) and substantiated by more recent studies (Haigh 2007; Lean & Rind 2008; Gray *et al*. 2010; Ineson *et al*. 2011; Ermolli *et al*. 2013; Solanki *et al*. 2013).

Total solar irradiance (TSI), the spatially- and spectrally-integrated radiant energy from the Sun incident at the top of the Earth's atmosphere and normalized to one astronomical unit, has been measured with space-borne instruments continuously since 1978 (see Figure 1). This measure averages 1361 W m⁻² (Kopp & Lean 2011) with typical increases of ~0.1 % from the minimum to the maximum of the 11-year solar cycle during recent decades (Fröhlich 2006). Additional and occasionally larger variations occur as sunspots and facular magnetic features emerge, transit, and decay on the Earth-facing portion of the solar disk.

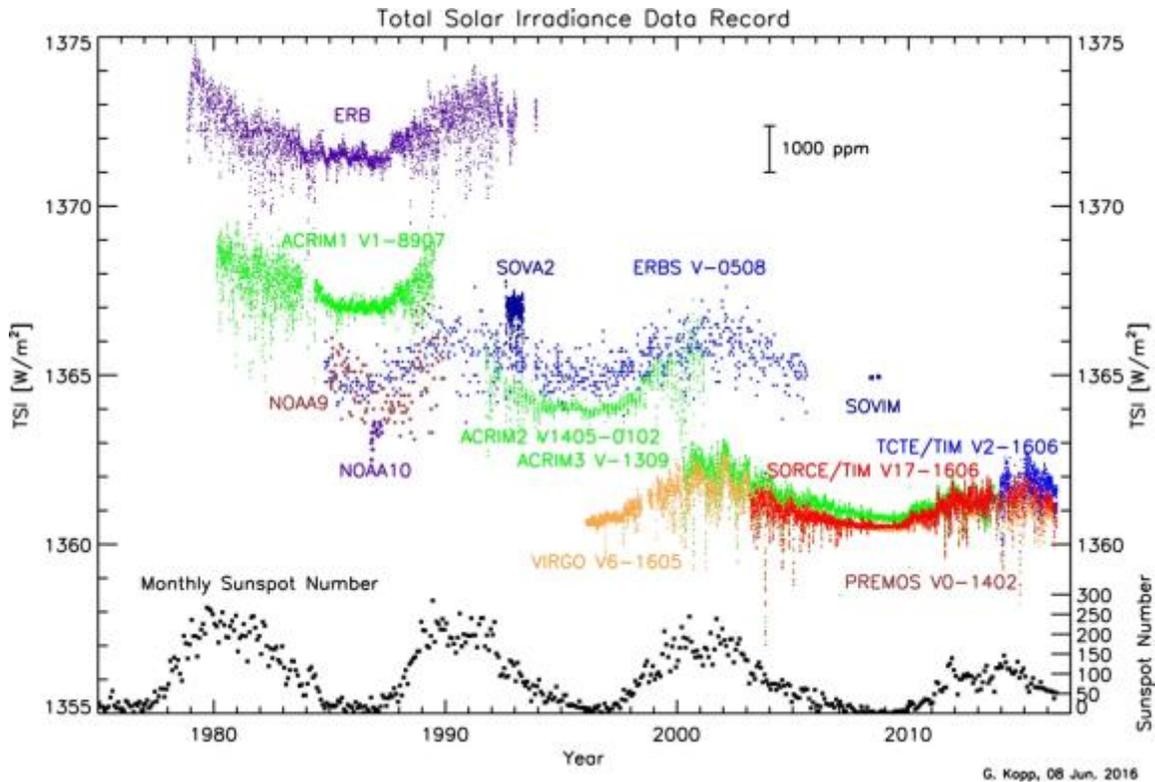

Figure 1: The TSI has been measured from space via an uninterrupted series of overlapping instruments since 1978. Offsets between instruments are due to calibration differences. Note that TSI fluctuations of approximately 0.1 % are in phase with activity over the 11-year solar cycle, as indicated by sunspot numbers.

Overlap between successive instruments enables the creation of composite records of solar variability spanning this space-borne measurement era and largely accounts for offsets and trends between different instruments. Three prominent composites are produced by different instrument principal investigators (Fröhlich 2006; Willson & Mordvinov 2003; Dewitte *et al*. 2004). While these composites generally agree on short-term variations in the Sun's output, weightings of and corrections applied to the individual instruments included in each composite cause different long-term trends between the three, as not all instruments have equal on-orbit stability. Fröhlich's (2006) Physikalisch-Meteorologisches Observatorium Davos (PMOD) composite, shown in Figure 2, includes several corrections for suspected instrument artifacts affecting the earlier instruments, particularly those affecting the NIMBUS7/ERB as discussed by Chapman *et al*. (1996) and Lee *et al*. (1995). Showing reasonable consistency between the TSI record and independent indicators of solar variability, this is generally considered to be the most solar-representative composite, as evidenced by its selection in the 2013 Intergovernmental Panel on Climate Change (IPCC) Fifth Assessment Report (AR5:Myhre *et al*. 2013), and best matches the two most prominent solar-irradiance reconstruction models (Krivova, Vieira, and Solanki 2010; Lean 2000); although even this composite has known issues limiting secular-trend detection, as discussed in §2.2.1.

In addition to the offsets between individual instrument measurements in Figure 1, each has different trends with time due to applied corrections for on-orbit degradation, thermal effects, and instrument anomalies. The instruments are ambient-temperature electrical substitution radiometers in which an absorptive cavity located behind a precision aperture measures incident radiant power. The ratio of sunlight power absorbed to the aperture area gives solar irradiance. Corrections are applied to account for the cavities' efficiencies based on pre-launch ground calibrations and for on-orbit degradation of the absorptive cavity-interior surfaces due to long-term solar exposure, which includes unfiltered ultraviolet light and x-rays. This on-orbit degradation is tracked via inter-comparisons between a primary cavity used for nearly continual solar monitoring and lesser-used cavities in each instrument that provide redundancy and tracking of changes in the primary due to solar exposure. First implemented by the ACRIM1 (Willson 1979), this degradation-tracking technique has been used by all subsequent TSI instruments. Measured net on-orbit degradation for current flight instruments ranges from 0.4 % for the SoHO/VIRGO and Picard/PREMOS to 0.02 % for the more stable SORCE/TIM, as shown in instrument assessments by Kopp (2014). The magnitude of these degradations and how well they can be tracked determine instrument measurement-stability uncertainties.

Measurement stability is critical for correlating solar variability with Earth climate records, which rely on multi-decadal- to millennial-length records of climate indicators, such as temperatures, sea-surface levels, glacial extents, and tree rings. Comparisons between these climate records and historical reconstructions of solar irradiance, such as provided by Unruh *et al*. (1999), Fligge and Solanki (2000), Krivova *et al*. (2003), Lean (2010), Ball *et al*. (2011), and Coddington *et al*. (2015), enable attribution of climate effects to their influences. Regressions of natural forcings, including solar variability and volcanic eruptions, and anthropogenic forcings, such as varying greenhouse-gas emissions, suggest that solar variability accounts for less than 10 % of climate change over the last century (Lean 2010). The more recent 2013 IPCC AR5 (Myhre *et al*. 2013), placing a range on TSI radiative forcing over the period 1745 to 2008 of 0.0 to 0.10 W m$^{-2}$, suggests that solar influences are even smaller than in the previous IPCC AR4 (IPCC 2007), causing a nominal 2 % contribution (with a range from 0 to approximately 8 %) to climate change over this period. While dominated by anthropogenic forcing in these recent times, solar variability in prior eras caused much larger relative influences. Determining climate sensitivity in these pre-industrial times is needed for validations of global- and regional-climate models; and the historical solar irradiance reconstructions needed for such Sun/climate comparisons all rely on knowledge of solar variations during the space-borne TSI measurement record.

## 2. Total Solar Irradiance Variability

The most accurate measurements of TSI variability are from space-borne instruments, as the Earth's highly-variable atmospheric transmission precludes ground-based measurements having capabilities to detect typical <0.1 % solar irradiance variations over days to years. Correlating measured TSI-variability with solar features, such as surface manifestations of magnetic activity as described by Domingo *et al*. (2009), enables longer-term measurements of those features to estimate solar irradiances prior to the direct measurement record, extending the knowledge of solar variability beyond the current 37-year space-borne record duration.

## 2.1. Solar Variability Within the TSI Measurement Record

### 2.1.1. Variability on Solar-Rotational Timescales

On 27-day solar-rotation timescales, the formation and disappearance of sunspots and faculae account for nearly all TSI variability (Lean 2010). Willson and Hudson (1991) report that sunspots account for the majority of the variability on timescales of a few days while faculae account for most of the variability over a few weeks. The passage of sunspots across the Earth-facing portion of the solar disk causes short-term decreases in the TSI and is responsible for most of the high-frequency variations in the daily data shown in Figure 1, while more spatially-extended and longer-lasting faculae, magnetically-active regions which are generally associated with and surrounding sunspots, cause brightening over longer timescales. The passage of large sunspot groups across the disk in late October 2003 caused the largest short-term decrease in TSI ever recorded, dimming the Sun by 0.34 % (Kopp *et al.* 2005). Over longer durations, however, facular brightening dominates such that the TSI is greater at times of higher magnetic activity. Fröhlich and Lean (2004) and Lean (2010) determined that much of the observed short-term variability can be estimated by empirical TSI proxy models based on linear regressions to these two components alone. The more sophisticated semi-empirical SATIRE model (Solanki *et al.* 2005; Krivova *et al.* 2011; Ball *et al.* 2011; Yeo *et al.* 2014), which includes additional solar activity indicators combined with physical models to provide further refinements, obtains a similar fundamental conclusion – the irradiance variability on solar-rotation timescales is largely due to the opposing effects of sunspot darkening and facular brightening.

Solar variations on these timescales are too rapid to affect Earth's climate via direct forcing from TSI because of the large climate-system heat capacities involved. Indirect mechanisms due to spectral solar irradiance variations, as described by Haigh (2007), may lead to climate effects from solar-rotational variations influencing longer timescales by affecting Earth-atmospheric circulation patterns, however. Such mechanisms are beginning to be modeled, but the magnitudes of their effects are not yet known.

### 2.1.2. Variability on Solar-Cycle Timescales

On solar-cycle timescales the TSI varies by ~0.1 %, as first reported by Willson and Hudson (1991) using the full dataset from the SMM/ACRIM1 spanning the period 1980-1989. The TSI variability over subsequent solar cycles has been similar in magnitude with the variability being in phase with solar activity (as represented by the sunspot number in Figure 1). The PMOD composite shown in Figure 2 indicates peak-to-peak amplitude variations of 0.082 %, 0.077 %, 0.096 %, and 0.063 % respectively for the peaks of each of Solar Cycles 21 (peaking in 1980) through 24 using annual medians of daily measurements.

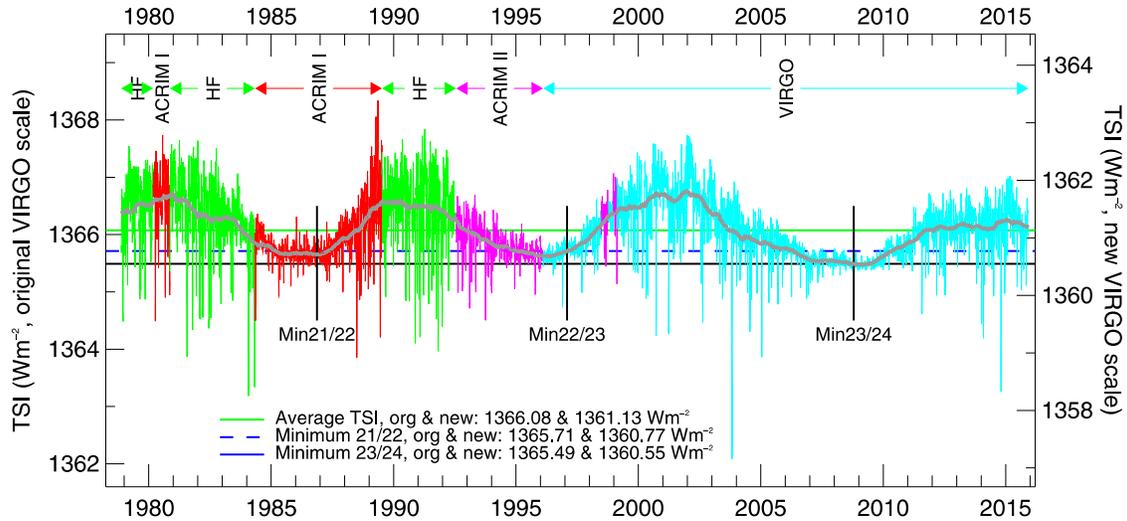

Figure 2: The PMOD TSI composite shows peak-to-peak TSI variability of ~0.1 % in each of the three solar cycles observed during the space-borne measurement record, with that variability being in phase with solar activity. The colors indicate the binary selections of different instruments used in the creation of the composite. The right-hand vertical scale indicates the more accurate currently-accepted absolute value. ("HF," short for its creators Hickey and Friedan, is a name used by some authors for the NIMBUS7/ERB instrument in Figure 1.) (Figure is courtesy of the VIRGO team)

On solar-cycle and solar-rotation timescales, the majority of the TSI-amplitude fluctuations are the result of opposing brightenings caused by faculae and shorter-duration dimmings caused by sunspots. Lean (2010) finds that these two manifestations of solar-surface magnetic activity explain 93 % of the TSI variability observed by the SORCE/TIM instrument and 83 % of that in a TSI composite over a solar cycle or longer. Chapman *et al.* (2012), using different solar proxies than Lean, obtain very similar results on these timescales, finding that two photometric sums obtained from ground-based observations explain 95 % of the SORCE/TIM variability over the first 7 years of its measurement record. Applying this empirical method over the same time range to the three TSI composites discussed in §2.2.1, these authors report matching the ACRIM composite to 88.7 %, the PMOD to 92.2 %, and the RMIB to 92.4 % (Chapman *et al.* 2013), indicating again that over solar-cycle timescales much of the variability in TSI can be accounted for via proxies of manifestations of solar-surface-magnetic features.

Variability on solar-cycle timescales is well established by the TSI measurement record and has been linked to similarly well-established Earth climate records, indicating a definitive Sun-climate connection. The typical 0.1 % solar forcing on these 11-year timescales can be seen in global- and regional-temperature records, sea-surface levels, ozone, tree rings, and precipitation amounts (Lean & Rind 2008; Lean 2010; Gray *et al.* 2010). Lean (2010) reports that globally-averaged surface temperatures increase about 0.1°C with irradiance increases in recent solar-cycles. Lean and Rind (2008) show regional-temperature changes linked to the solar cycle can be even larger on these timescales.

### 2.1.3. Variability on Minute- to Hour-Timescales

On timescales of minutes to hours, the TSI varies at the ~0.01 % level due to the globally-averaged superposition of solar convection and oscillations (Fröhlich & Lean 2004; Kopp *et al.* 2005). These ever-present variations in TSI have periods in the 3- to 10-minute range typical of their solar drivers. An example of such short-term variability is shown in Figure 3.

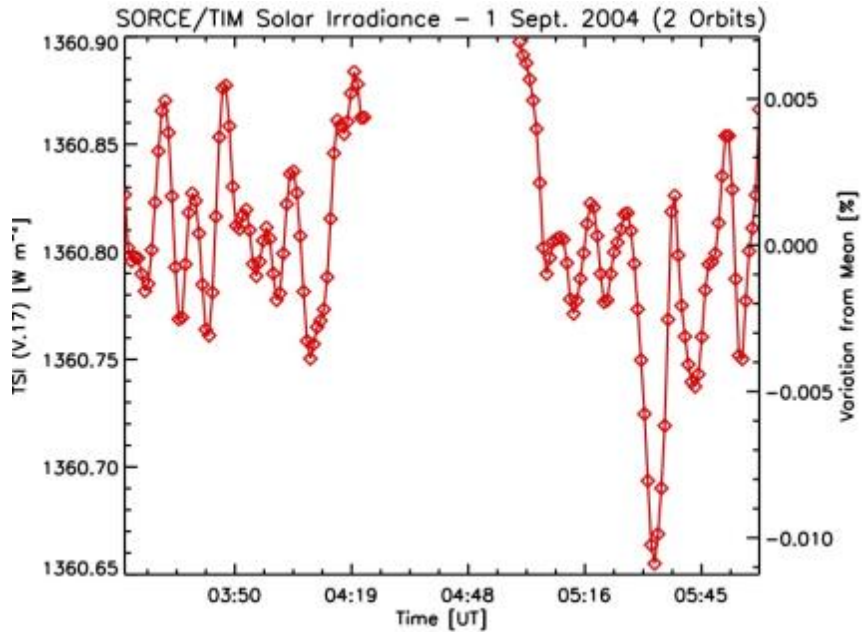

Figure 3: The TSI varies continually at the ~0.01 % level on timescales of several minutes due to the superposition of convection and oscillations (updated from Kopp *et al.*, 2005).

Solar flares have occasionally been observed in TSI (Woods *et al.* 2003, 2006). While flares are prominent in the ultraviolet and x-ray spectral regions, which have low background intensity from the quiet Sun, they are small in both spatial extent and in net energy compared to the total solar irradiance. The fourth largest flare ever recorded by the GOES X-Ray Spectrometer measurements, an X17 flare near disk center on 28 October 2003, caused an abrupt but short-duration 0.028 % increase in the TSI (Kopp *et al.* 2005). As shown in Figure 4, the abrupt initiation of this flare clearly exceeded the ever-present solar-background variations due to convection and oscillations.

By measuring the spectrally-integrated energy of the Sun, flare observations in TSI provide the net radiant energy released by the flare, which cannot be directly obtained from narrow-band spectral observations. Producing so little energy for such a short duration compared to the steady net radiation from the quiescent Sun, flares have no direct radiative effect on Earth's climate.

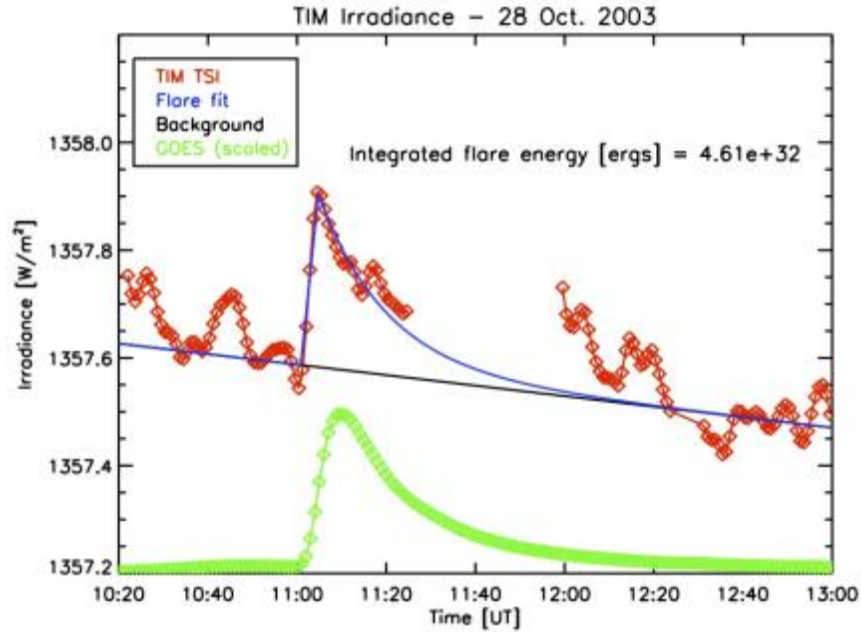

Figure 4: The X17 flare on 28 October 2003 caused an abrupt, short-duration 0.028 % increase in the TSI. A time-integration of a fit of the flare's abrupt increase and subsequent decay (blue) to TSI (red) shows a net radiant-energy release of 4.6x10²⁵ J for this flare. (from Kopp et al., 2005)

### 2.1.4. Effects of Sun-Like Variability on Detecting Exo-Solar Planets via Parent-Star Transits

Sun-like stars may similarly be expected to vary in integrated brightness continually at the 0.01 % level. Thus the ability to detect the transit of an exo-solar Earth-like planet in front of a parent Sun-like star may be difficult, as the dimming caused by such a transit is typically <0.01 %. However, while the Sun varies at this level on timescales of several minutes and shows larger variations over days to weeks, it shows little variability on the ~10-hour timescales typical of planetary transits across the disks of Sun-like stars from orbits in the habitable zone, where liquid water can exist on the planet's surface. High-cadence TSI data from NASA's low-noise SORCE/TIM instrument combined with a simulated transit of an Earth-sized planet showed that such transits were likely detectable despite the typical background solar-variability. Subsequently, the Kepler mission (Livio et al. 2008), launched in 2009, has discovered over 10 Earth-sized planets orbiting in the habitable zone using stellar photometric techniques to observe their transits. This mission has photometric capabilities marginally capable of detecting an Earth-type planet occulting a Sun-like star, so relies on at least three transit observations for discovery of an exo-solar planet. This photometric planet-detection technique favors finding large planets having close orbits (providing frequent transits) of small and stable stars.

Observations of inner-planet transits of the Sun from the Earth in total irradiance can indicate the photometric sensitivities needed for such transit detections that may be partially masked by normal solar variability. The SORCE/TIM TSI instrument has observed two transits each of Mercury and Venus (Kopp et al. 2005; Kopp & Ward 2012). Even though these are smaller planets than the Earth, from the SORCE's low-Earth-orbit vantage point they cause larger relative transit dimmings than they would if viewed in a distant exo-solar system. From the Earth's view, a transit of Venus causes about a 0.1 % transit depth while a Mercury transit causes a much smaller 0.004 % dimming. The Venus transits are, as expected, unmistakable with this TSI instrument's <0.001 % radiometric sensitivity. The Mercury transits, however, are largely hidden by the much greater background signals due to solar convection and oscillations, and cannot be considered unambiguous detections. Figure 5 shows the TSI variations during a transit of each of these inner planets as observed by the Earth-orbiting SORCE/TIM.

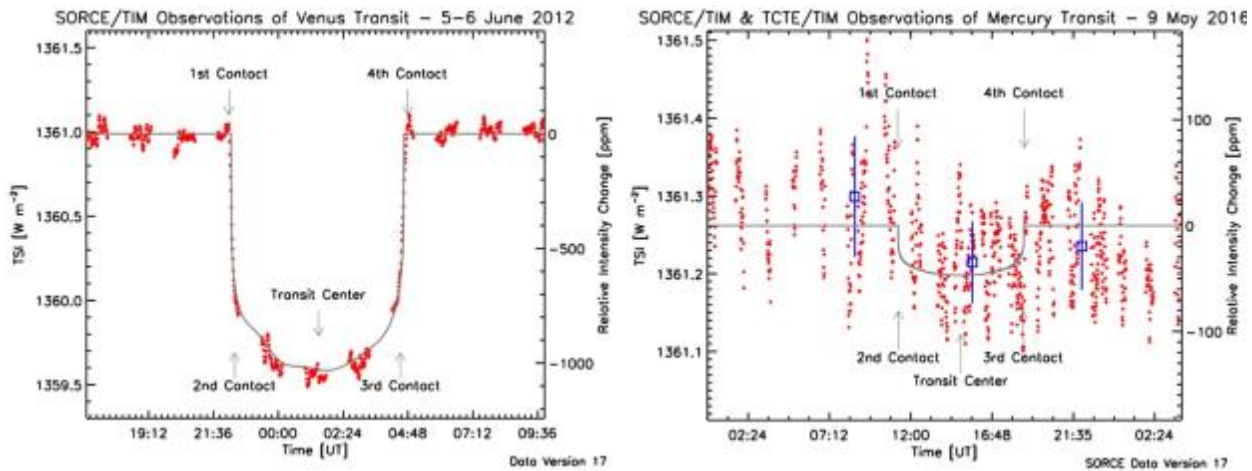

Figure 5: Transits across the solar disk for each of the inner planets, as measured in TSI by the Earth-orbiting SORCE/TIM and TCTE/TIM instruments, should have decreases of ~0.1 % for Venus (left-hand plot) and 0.005 % for Mercury (right-hand plot). The calculated TSI signal during each transit (grey) includes the effects of solar limb-darkening and the spacecrafts' orbital positions. The actual TSI measurements (red dots) include high-frequency variations due to solar convection and oscillations. While the 2012 Venus transit is readily apparent in TSI, this high-frequency background solar-variability is comparable to the Mercury-transit signal. Averages of the TSI values from equal-duration times before, during, and after the 2016 Mercury transit (blue squares) indicate a measured decrease of 0.004 %. While this transit is suggestively discernable in the red TSI data despite the underlying continual solar variations, the uncertainties on this decrease are comparable to the value itself.

## 2.2. Solar Variability on Secular Timescales

While climate signatures due to TSI variability on solar-cycle timescales are well established, longer-term climate influences from the Sun are less well quantified; yet such long-term changes can be even more important, as secular variations in the Sun's energy can have more direct influences on Earth's climate, particularly in pre-industrial times when natural influences dominated climate forcings. Sensitivities to natural forcings can be determined, for example, by correlating historical solar-irradiance variability estimates with observational surface-temperature records. Reliable knowledge of possible secular changes in TSI during these earlier eras is therefore important for establishing climate sensitivity to natural influences in the absence of larger anthropogenic effects.

### 2.2.1. Secular Variability over the Space-borne Measurement Era

There are two measurement techniques for determining long-term trends: 1) Acquire continual measurements using instruments having stability uncertainties that are lower than the trends to be detected; or 2) obtain measurements over a sufficiently-long duration that the trend exceeds the absolute accuracy of potentially-disparate instruments acquiring the measurements, such that continuity is not required. The latter approach, which has advantages in detecting continual trends over long periods of time, requires TSI-measurement uncertainties of ~0.01 %; the former approach requires measurement continuity plus instrument-stability uncertainties of less than 0.001 % $yr^{-1}$ from every instrument contributing to the continuous record, and is better able to detect large short-term variations (Ohring *et al.* 2007; Kopp 2014). As an example of this transition between the favored approaches with measurement duration, a potentially-realistic TSI-variability trend as large as 0.1 % over 100 years, such as discussed in §2.2.2, may be detected at the 1-$\sigma$ level by continual and overlapping instruments having 0.001 % $yr^{-1}$ stability uncertainties even over short time periods (but regardless of measurement duration), while non-overlapping instruments having 0.01 % uncertainties on an absolute scale could provide a similarly-marginal 1-$\sigma$ level trend-detection only after 10 years (but would improve on detection of an assumed-continual trend for greater durations).

TSI variability on secular timescales is thus not definitively known from the space-borne measurements because this record does not span the desired multi-decadal to -centennial time range with the needed absolute accuracies, and composites based on the measurements are ambiguous over the time range they do cover due to high stability-uncertainties of the contributing instruments.

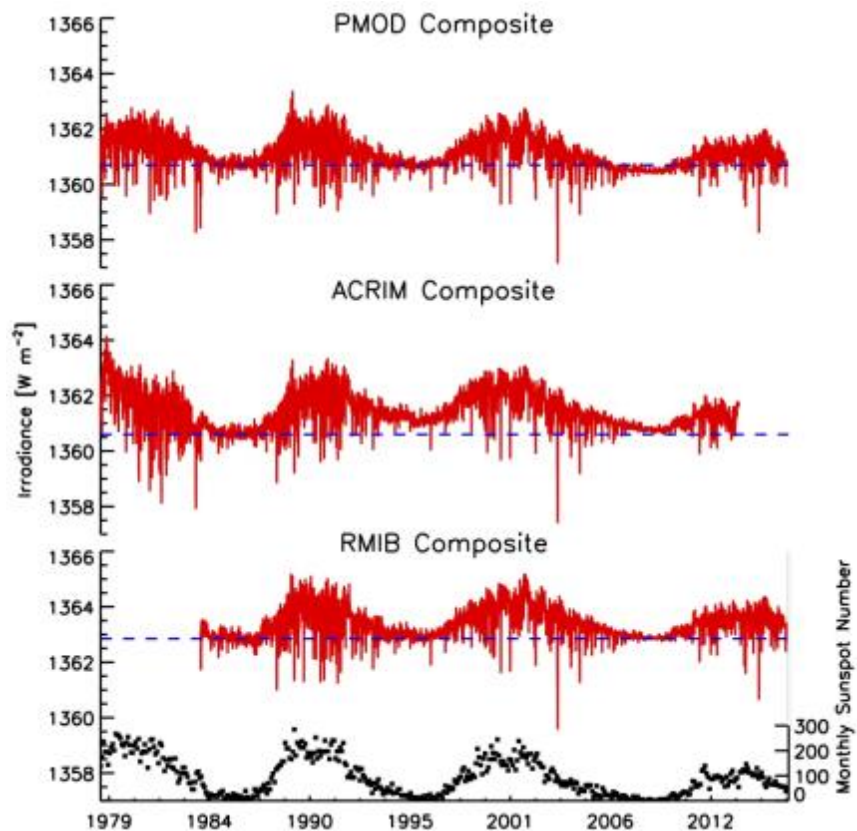

Figure 6: The three most prominent TSI composites show different trends based on the selection of how each contributing instrument is weighted or corrected. These uncertainties limit the ability to definitively discern a secular trend over the measurement record.

The three most prominent TSI composites, plotted in Figure 6, show differences in trends between solar minima that are suggestive of secular solar-variability. Not all composite's trends can be correct, however, since they are all representations of the same Sun. Fröhlich (2009) reports a decrease in the PMOD composite of 0.0012 ± 0.0008 % yr$^{-1}$ between the two most recent solar minima (occurring in 1996 and 2008). Willson and Mordvinov's (2003) ACRIM composite indicates a 0.005 % yr$^{-1}$ increase between the 1986 and 1996 minima followed by a comparable decrease to the subsequent minimum. Their reported increase between the earlier two minima led to claims that solar variability was responsible for as much as 69 % of the global temperature increase over the last century (Scafetta & West 2008), although that work was quickly discredited (Foukal 2008; Schmidt 2008; Lean 2010). Since each composite is based on a unique, time-dependent selection of which instrument is used for which time ranges, differences between the composites are one indicator of the level of instrument instabilities.

Comparisons to solar models provide another indication of limitations in TSI instrument-stability uncertainty. Extending their two-component empirical-model mentioned in §2.1.1 to longer-term composites, Chapman *et al.* (2013) report matching the three composites in Figure 6 to 73.7 % for ACRIM, 87.7 % for PMOD, and 84.5 % for RMIB. The fits to these composites, which span 21.5 years via multiple instruments, is poorer than fitting to an individual instrument, as in §2.1.1, partially because of stability- and trend-differences between the multiple instruments contributing to the composites. A similar time-range comparison using the more sophisticated semi-empirical SATIRE model, presented in §2.2.2, replicates 97% of the SORCE/TIM TSI variability over the period from Feb. 2003 through Oct. 2009 (Ball *et al*. 2011), 96% of the variability in the strictly-VIRGO-based portion of the PMOD composite from 1996 to 2013, and 92% of the PMOD composite over the entire period from 1978 to 2013; although a portion of that poorer agreement at these earlier times is likely attributable to the lower quality of the solar magnetograms upon which SATIRE is based then (Yeo *et al*. 2014). These poorer model-to-measurement fits over lengthier time ranges compared to the shorter-duration records is another indication of the likely limited measurement-stability during the space-borne era.

Yet a third indicator of stability uncertainty in an instrument is the variation in that instrument's data from one released data version to another. While updated data versions are intended to include improvements based on new knowledge of the instrument's status, version-to-version changes indicate that the former data version was inaccurate by (at least) the difference between the two. The August, October, and December 2015 VIRGO data releases provide an example, whereby successive pairs of these three releases showed trend differences of -0.0004 % yr$^{-1}$ and +0.0006 % yr$^{-1}$, lending doubt that the actual VIRGO instrument stability is known to better than these levels (see Figure 7). These data releases commensurately affect the PMOD composite, which is based primarily on the VIRGO data. Note that these version-to-version variations are nearly as large as the PMOD composite's reported solar minima-to-minima trends mentioned previously. Even greater trend-changes were included between successive versions of ACRIM3 data in 2011, when erroneous ~0.02 % peak-to-peak cyclic TSI-variations caused by annual fluctuations in instrument temperatures were corrected. While such corrections presumably improve the later releases, they do indicate the levels of errors in previous releases and may call into question whether all needed corrections are truly included in the latest release.

Considerations such as these in addition to the disparate composites proposed by different researchers indicate the limitations of the TSI-record stability level. This assessment is not intended to detract from the usefulness of the TSI-measurement record, but to point out to users of the TSI composites the limitations of those records, as none of the existing composites include time-dependent uncertainties to indicate their long-term trend-detection capabilities.

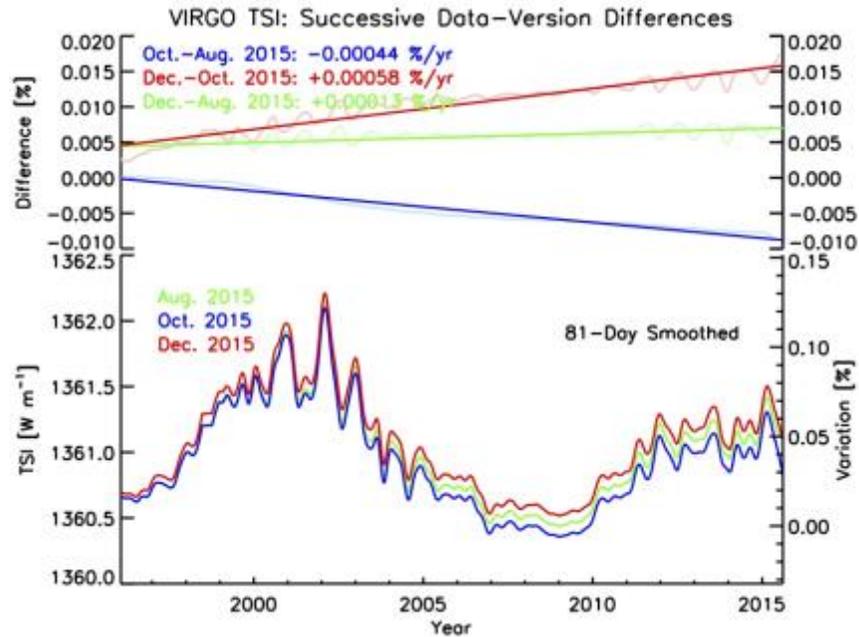

Figure 7: Successive releases of VIRGO TSI data separated by two months each (lower plot) show significant trend differences (upper plot) due to variations in data-processing and estimates of instrument degradation with time.

With large activity-caused variations between maxima in different solar cycles, possible secular trends in TSI variability may be most evidenced by changes between solar-minima values. While some TSI composites do show such trends, such as a possible decrease between the 1996 and 2008 solar minima, they are inconsistent and not supported by likely instrument-stability uncertainties. For context, however, other indicators of solar variability similarly show a downward trend between these last two minima. Solomon *et al*. (2011) attribute the Earth's lower thermospheric density during the 2008 minimum, as evidenced by drag measurements of Earth-orbiting spacecraft, to be due to lower levels of extreme-ultraviolet solar-irradiance. Similarly, Pacini and Usoskin (2015) use neutron-monitor records from eight Earth-based stations to show an increase in polar cosmic-ray intensities during the 2008 solar minimum from the prior three, which they attribute to changes in the solar wind. While such independent observations are indicative of lower solar activity during this recent solar minimum and may be corroborative of possibly-similar secular trends in TSI-variations during the 2008 solar minimum. However, the TSI is not necessarily correlated with these solar-variability indicators, so they should not be perceived as justification for the claimed magnitudes of variations between solar minima shown in the composites in Figure 6.

The TSI-record assessment in this section suggests that the current measurement-record does not have the needed stability to definitively detect secular trends in solar variability. Improvements to instrument stability and/or absolute accuracy are needed to detect secular variations at the <0.001 % yr$^{-1}$ level required for secular solar-variability detection and climate studies (Kopp 2014).

### 2.2.2. Secular Variability on Climate Time Scales

TSI models provide a means of extending the measurement record to historical times via reconstructions based on indicators of solar activity related to TSI variability.

Solar-irradiance variability can be modeled by correlating direct TSI observations with indices representative of solar-surface magnetic features, the most prominent of which are sunspots and faculae. Solar irradiance is then constructed at prior times by applying the correlations to historical indices of these solar features. A prominent "empirical" solar-proxy model is the Naval Research Laboratory Total Solar Irradiance (NRLTSI: Lean 2000; Lean *et al*. 2005; Coddington *et al*. 2015). This model provided the historical solar-irradiance records used in both the IPCC's AR4 and AR5 from 1610 onward when sunspot records were available (IPCC 2007; Myhre *et al*. 2013). In the so-called "semi-empirical" class of models, solar magnetic-feature observations or proxies are applied less directly by assessing the solar-disk-area coverage from spots and faculae and then weighting the time-independent brightness contrasts of the corresponding components computed from models of the solar atmosphere. An example of a TSI-reconstruction model of this type is the Spectral And Total Irradiance Reconstructions model (SATIRE: Krivova *et al*. 2011; Dasi-Espuig *et al*. 2014; Yeo *et al*. 2014). An unweighted average of this model and an updated version of the NRLTSI model (NRLTSI2) used in the previous two assessments will provide estimated solar influences in the upcoming Coupled Model Inter-comparison Project Phase 6 (CMIP6) for the next IPCC report.

These and several other TSI reconstructions are shown in Figure 8. All rely on correlations with the current spacecraft-era measurement record but include different methods of accounting for possible long-term variability not measured directly by the record. Lean (2000) uses a combination of the group-sunspot-number and an 11-year running mean along with distributions of cyclic and noncyclic Sun-like stars to estimate secular solar-variability. Wang *et al.* (2005) improve on this via a solar magnetic-flux-transport model fitted to variations in the observed open solar magnetic-flux, making their reconstruction solar-specific and not dependent on stellar variabilities. Tapping *et al*. (2007, not plotted) correlate variability with $F_{10.7}$ radio emission over the TSI measurement record and then extend this record back in time directly via the 10.7-cm measurements to 1947 and prior via similar correlations to sunspot records.

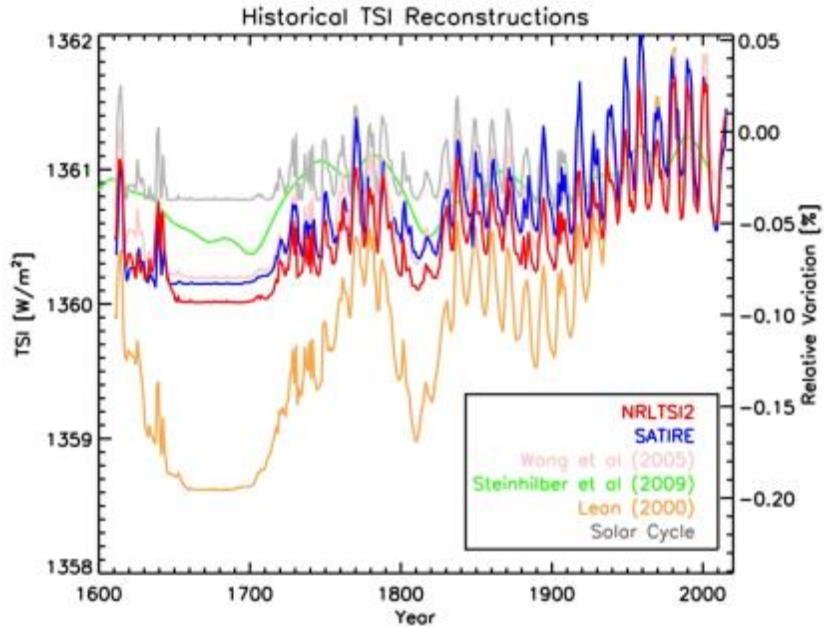

Figure 8: TSI-reconstruction models provide estimates of solar variability from the Maunder Minimum to the present.

TSI reconstructions over the time range in Figure 8 rely on the sunspot record, which began in 1610. Most use the group-sunspot-number of Hoyt and Schatten (1998). This record has recently come under scrutiny, with Clette *et al.* (2015) suggesting a very different reconstruction based on their reanalysis of the records from individual observers contributing to the long-term sunspot record. Using their "backbone" method of linearly-regressing sunspot counts from each observer throughout the period, they reconstruct the sunspot record back to 1700. Lockwood *et al.* (2016) question the use of this method because of its potential to propagate errors in time, the limitations of linear regressions in this application, and the assumption that each observer maintains a time-independent sensitivity in their counting of sunspots. These authors instead favor a statistical approach by Usoskin *et al.* (2016), which does not require such assumptions and produces a sunspot record more similar to that of Hoyt and Schatten. Kopp *et al.* (2016) estimate the effects the resulting sunspot records from these two camps have on the NRLTSI2 and the SATIRE TSI-reconstruction models.

Other solar-reconstruction models extend the TSI record to times before the sunspot record by using cosmogenic isotopes (Vieira *et al.* 2011; Delaygue & Bard 2011; Steinhilber *et al.* 2012; Usoskin 2013). These authors reconstruct historical TSI for thousands of years by correlating solar variability estimated over the 400-year sunspot-measurement record to cosmogenic-isotope depositions determined from ice cores and tree rings, as these depositions are influenced by open solar magnetic-flux in the heliosphere. The near-term portion of one such reconstruction by Steinhilber *et al.* (2009) is included in Figure 8.

The more recent of the reconstructions shown in Figure 8 indicate that TSI values during the Maunder Minimum, a period from 1645 to 1715 having very few sunspots and likely low solar activity (Eddy 1976), were 0.5 to 1 W m⁻² (0.04 to 0.08 %) lower than the mean present-day values. Judge *et al*. 2012 claim a more extreme variation having Maunder Minimum values lower by 3 W m⁻² (0.2 %). This larger variation cannot be excluded based on current knowledge of correlations between proxies available during that period and the TSI itself.

These long-term estimates of solar variability are needed for climate studies extending into the pre-industrial era. Gray *et al.* (2010) provide a review of solar effects on climate and the atmosphere on solar-cycle and longer timescales. Determining climate influences to historical solar-forcings are limited not only by uncertainties in TSI reconstructions but also by those in Earth's climate records, which similarly rely on proxies for historical reconstructions.

In conjunction with climate effects due to solar variability on secular timescales, variations in Earth's orbital parameters change the Sun's energy incident at the Earth and provide additional radiative forcings on 19 000, 23 000, 42 000, and 100 000-year timescales (Hays *et al.* 1976). These Milankovitch-induced variations on the solar energy reaching the Earth manifest primarily in long-term latitudinally-dependent insolation variations and globally-averaged seasonal effects. While important for historical climate studies on long-term timescales, these orbit-parameter effects are only mentioned here for their climate relevance but are not described more fully as they are not intrinsic variations in the Sun's radiative output.

### 2.2.3. Variability Due to Solar Evolution

On billion-year timescales, the Sun changes by much larger amounts than at the (by comparison relatively short) timescales discussed above. Currently a main-sequence star on the Hertzsprung–Russell diagram, the Sun varies due to stellar evolution as it burns its hydrogen supply along this sequence expected to last a total of roughly ten-billion years. The early Sun was approximately 70 % as bright as at the present when it joined the main sequence about 4.6 billion years ago and has a current rate of increase in luminosity of 0.009 % per million years (Hecht 1994). At this rate, it will take ten-million years for the background solar-brightness to increase by the 0.1 % typical of a solar-cycle variation, and another 3.5 billion years for heating from the Sun to create Earth-surface conditions similar to those of the present-day Venus; although additional effects, such as feedback from enhanced ocean-evaporation, may accelerate this warming and make the Earth uninhabitable (at least to present-day complex lifeforms) in about one-billion years (O'Malley-James *et al.* 2013). In 4.5 billion years or so, the Sun will leave the main sequence and transition to a red giant, increasing in radius by about 250 times and in luminosity by roughly 27 000 times but decreasing in temperature to 2600 K, spectrally shifting its radiant output toward longer wavelengths. Such spectral shifts occur even prior to leaving the main sequence, as increases in solar luminosity, radius, and temperature along this sequence cause both increases in the total solar irradiance as well as shifts of the spectral solar irradiance toward shorter wavelengths with time.

Hecht (1994) gives good technical details of this evolution while Eddy (2009), in his usual colorful and entertaining style, gives a non-technical overview of the Sun's evolution and future.

## 3. Discussion and Conclusions

The TSI has been monitored continually from space-borne instruments since 1978, providing the most accurate measurements of the net energy driving Earth's climate system. These measurements show variations in the TSI at the 0.01 % level on timescales of a few minutes due to solar convection and oscillations. A few extremely large flares have been observed to cause larger TSI increases on these timescales. Over days to weeks, surface manifestations of solar magnetic activity cause changes at the ~0.1 % level. These are largely due to the opposing effects from sunspots, which cause decreases in TSI when near disk center, and faculae, which cause a net brightening. Over solar-rotation to solar-cycle timescales, the facular brightenings offset the sunspot darkenings such that the TSI is about 0.1 % brighter during times of solar maximum.

Less is definitively known about variations on secular timescales. TSI reconstructions based on historical indicators of solar activity that are correlated with the current measurement-record provide estimates of past variability. Generally-accepted results indicate that the Sun may have been about 0.04 to 0.08 % lower in irradiance for several decades during the Maunder Minimum in the late seventeenth century than current levels. Yet longer timescales, using cosmogenic isotopes as solar-activity indicators, suggest similar levels of historical solar variability.

Limiting factors in determining long-term variability are both the current duration of the space-borne measurement-record and the contributing instruments' stability uncertainties. While these instruments provide some of the most stable measurements of any on-orbit radiometric devices, they are attempting to measure very small changes in the Sun's output radiant energy over very long time spans while being exposed to damaging ultraviolet and x-ray radiation. Instrument stability at the needed <0.001 % $yr^{-1}$ level has not been achieved by all instruments contributing to the current record, limiting the ability to create unambiguous composites spanning the entire 37-year record. Lower instrument-stability uncertainties via more robust instruments having less net degradation combined with continued uninterrupted measurements should improve this solar climate data record in the future. Improvements to absolute accuracy will make the record less reliant on measurement continuity. These improved TSI measurements will refine solar models, which in turn will improve historical reconstructions and enable better climate-sensitivity estimates to solar forcing.

## Acknowledgements

This paper benefitted from insightful and broadening suggestions from reviewer Dr. W. Ball and from a careful reading by an anonymous reviewer. Additionally I gratefully acknowledge the support of NASA's SORCE (NAS5-97045) and SIST (NNX15AI51G) for this effort. Figure 1 includes data from: www.ngdc.noaa.gov/stp/SOLAR/solar.html (NIMBUS7/ERB, ERBS/ERBE, NOAA9, and NOAA10); http://www.acrim.com (ACRIM1, ACRIM2, and ACRIM3); the VIRGO team via ftp://ftp.pmodwrc.ch; http://lasp.colorado.edu/home/sorce/data/tsi-data/ (SORCE/TIM); the PICARD/PREMOS team (personal communication, A. Fehlmann, 2014); and http://lasp.colorado.edu/home/tcte/data/ (TCTE/TIM). Figure 2 is courtesy of the VIRGO team via ftp://ftp.pmodwrc.ch/pub/Claus/ISSI_WS2005/ISSI2005a_CF.pdf and explained by Fröhlich (2006). Figure 6 uses composite data from these sites as well as from ftp://gerb.oma.be/steven. The editor thanks William Ball and an anonymous referee for their assistance in evaluating this paper.